\documentstyle[12pt]{article}

\def\appendix{\setcounter{equation}{0}
\def\theequation{A.\arabic{equation}}}

\title{ \rm \begin{flushright}
            {\normalsize hep-ph/9503405} \vspace*{-2mm} \\
            {\normalsize IHEP 95-1} \hspace*{7.5mm} \vspace*{-3mm} \\
            {\small (Revised version)} \hspace*{0.5mm}
            \end{flushright}  \vspace*{2cm}
        Pseudoscalar glueball and $\eta'$-meson \\
        in low-energy QCD expansion    }

\author{  M. L. Nekrasov \thanks{E-mail: nekrasov@mx.ihep.su}
          \medskip \\
          {\small\it Institute for High Energy Physics,
                            Protvino, 142284, Russia}  }
\date{}

\begin{document}
\maketitle

\begin{abstract}
\noindent
An effective chiral lagrangian of order $p^2$, describing the
interaction of light pseudoscalar (PS) mesons with $\eta'$-meson
and PS-glueball, has been determined taking into consideration
the renorm-group requirements imposed by QCD renormalization.
It is shown that the interpolating fields for the lowest singlet
quarkic and gluonic states, $\eta^0$ and $\eta^G$, may be
involved into the effective theory to be renorm-invariant
objects not mixing due to QCD renormalization. It is established
that the potential describing the ``mass'' term of the
lagrangian does not depend on $\eta^0$.  The dependence on
$\eta^G$ is permitted only when there is not direct interaction
between $\eta^0$ and $\eta^G$ out of the ``mass'' term without
the octet fields contribution. The peculiarities distinguishing
the glueball from excitation over $\eta^0$ have been considered.
\end{abstract}

\newpage

\section{Introduction}

According to well settled notions, a prediction for glueballs
is one of the brightest consequences of QCD. However, so far
there is no satisfactory solution for the problem of their
quantitative description. A certain reason is lack of clear
understanding of the point how one can separate the gluonic
contributions from the singlet quark ones in a model-independent
way. The correct solution of the problem encounters a set of
difficulties, in the long run connected with necessity to
observe the local gauge invariance of the theory. One important
aspect of the problem is the dependence of the quark-gluon
mixing on the scale of the UV renormalization.

The latter problem, as a rule, is not taken into consideration.
However, when investigating the wave functions of singlet mesons
it exhibits itself inevitably while the quarkic and gluonic
composite operators, generating these states, are mixed due to
the UV renormalization. For the pseudoscalar (PS) meson channel
this phenomenon was first described in \cite{E-T}. The further
investigation of the problem was carried out in \cite{N,S-V}.

The presence of the nontrivial UV renormalization gives certain
difficulties in generalization of PCAC to the case of
$\eta'$-meson. Really, Refs. \cite{S-V,K-P} have recently shown
that the straightforward generalization of the well-known PCAC
formula for $\pi^0 \to \gamma\gamma$ to $\eta' \to \gamma\gamma$
is inconsistent with the renorm-group, and therefore incorrect
in principle. The right formula for $\eta' \to \gamma\gamma$
involves a new renorm-invariant constant instead of the decay
constant of the axial quark current.  Moreover, it is claimed in
\cite{S-V} that one more term should be added to the right
formula for $\eta' \to \gamma\gamma$, describing the coupling of
the ``glue'' component of $\eta'$ to photons.

The present paper considers the problem of $\eta'$ together with
the problem of PS-glueball, because both states most likely are
mixing.  The investigation is carried out in the approach of the
low-energy expansion of QCD, which is also known as the chiral
perturbation theory. Earlier, this very approach allowed one to
describe the octet of light PS mesons $\pi, K, \eta$ of the
Goldstone nature (see, e.g., \cite{G-L-1,G-L-2}). In
\cite{E-G-P-DR} it was applied to describe the interaction
between the lightest PS mesons and heavier meson resonanses.
Analogously, the $\eta'$-meson and PS-glueball may be involved
into the effective chiral theory.  But the involving should be
performed providing for the renorm-group properties inspired by
QCD renormalization.

The next section of the paper presents the short review of the
necessary knowledge on the renorm-group properties of the
generalization functional of QCD in the presence of the
composite operators generating the singlet states of
$\eta'$-meson and a PS-glueball. In section 3 the effective
chiral lagrangian of order $p^2$ is determined, involving the
singlet interpolating field $\eta^0$ of the quarkic nature and
some additional singlet field, which may describe a PS-glueball
or an excitation over $\eta^0$.  Section 4 is devoted to the
detailed study of the general properties of the effective
theory. The relationship between the currents of the effective
theory and the composite operators of QCD is discussed. As well,
the restriction on the potential describing the ``mass'' term of
the lagrangian is obtained and the consequences of the
restriction are explored. The typical difference was found to be
between the contributions of the PS-glueball and the excitation
over $\eta^0$ into the effective chiral lagrangian.  Section 5
investigates the spectrum of the theory. Section 6 summarizes
and discusses the results of the paper.

\section{Chiral symmetry and UV renormalization for composite
         operators in QCD}

The effective chiral lagrangian for light PS mesons was most
consistently described in the approach of Gasser and Leutwyler
\cite{G-L-1,G-L-2} where its connection with the generalization
functional of QCD was preserved. The latter in the presence of
the composite operators, generating the octet of mesons and the
singlet quarkic and gluonic states, is
\begin{equation}
e^{iW(V,A,S,P,\Theta)} = \int {\cal D}\left[q,\bar q,G_{\mu}\right]\>
e^{i\int d^4 x {\cal L}_{QCD}(q,\bar q,G_{\mu};V,A,S,P,\Theta)}.
\end{equation}
Here ${\cal D}\left[q,\bar q,G_{\mu}\right]$ is the measure of
the functional integral, $V,A,S,P$ are the sources for the
related quarkic composite operators and their chiral partners,
$\Theta$ is the source for the operator of axial gluon anomaly,
usually regarded as the generator for the gluonic state. ${\cal
L}_{QCD}$ in (1) is the unrenormalized lagrangian of QCD with
the sources (the notations are obvious):
\begin{eqnarray}
& &{\cal L}_{QCD} = {\cal L}^0_{QCD} + \bar q\,\gamma_{\mu}\!
            \left(V_{\mu} \! + \! \gamma_{5} A_{\mu}\right) q -
            \bar q\left(S \! + \! i\gamma_{5}P\right) q
           + \Theta Q,						\\[1mm]
& &V = \sum_{a=0,1,\dots8}\left(\lambda^a/2\right)\,V^a,\quad \dots \qquad
		(\lambda^0=\sqrt{2/3}\,\mbox{\bf I}),        \nonumber \\
& &Q  = \sqrt{6} \, \frac{\alpha _{s}}{8\pi }\,\frac{1}{2}\,
\epsilon _{\mu \nu \lambda \rho }G^{A\mu \nu }G^{A\lambda \rho }.
\end{eqnarray}
Here the nonstandard multiplier $\sqrt{6}$ ($6=2N_f$) has been
introduced in the definition of $Q$, providing convenient
reading for the subsequent formulae. Without sources one should
set in (2) $S = {\rm diag}(m_u,m_d,m_s)$, $P = V = A = \Theta =
0$.

Lagrangian (2) exhibits a property of the local $U(3)_{L}\times
U(3)_{R}$ chiral invariance, provided that the quark
transformations
\begin{equation}
q_L \to \Omega_L(x)\,q_L, \qquad  q_R \to \Omega_R(x)\,q_R,
\end{equation}
$q_{L,R} = (1 \mp \gamma_5)/2\,q$, are accompanied by
compensating transformations of the sources:
\begin{eqnarray}
L_{\mu} & = & V_{\mu} - A_{\mu}
	     \to \> \Omega_L\,L_{\mu}\,\Omega_L^{\dagger} +
                i\,\Omega_L\partial_{\mu}\Omega_L^{\dagger},   \nonumber\\
R_{\mu} & = & V_{\mu} + A_{\mu}
	     \to \> \Omega_R\,R_{\mu}\,\Omega_R^{\dagger} +
             i\,\Omega_R\partial_{\mu}\Omega_R^{\dagger},      \\
M & = & \, S + iP \,\> \to \> \Omega_L\,M\,\Omega_R^{\dagger}. \nonumber
\end{eqnarray}
It is well known that in quantum theory the anomaly breaks the
lagrangian invariance.  Nevertheless, the symmetry may be
restored if one imposes the additional conditions on the
sources.  So, one can demand the rotation of $\Theta$
accompanied to $U(1)_{A}$-rotation of the quark fields:
\begin{equation}
\Theta \to \, \Theta + i\sqrt{1/6}\,\ln\,\det(\Omega_L\Omega_R^{\dagger})
	     = \Theta - \omega^0_5.
\end{equation}
Here $\omega^0_5 = (\omega^0_R - \omega^0_L)/2$ is the parameter
of $U(1)_{A}$-rotation. Condition (6) compensates completely the
effect of the gluon anomaly. To compensate the anomaly depending
on the external fields (the sources for the composite operators)
one needs an additional term. It is clear that it should equal
the Wess-Zumino term with opposite sign, constructed over the
nonet of some auxiliary external PS fields (auxiliary PS
sources). Note, owing to the fact that this additional term does
not depend on the dynamical fields of the theory, which are the
functional integral variables in (1), it does not change the
dynamical properties of the theory. In particular, it does not
change its UV behaviour.

On the contrary, the insertion of the source-terms in (2) means
that the new kinds of interaction are introduced into the
theory. They may produce new kinds of UV divergencies. In order
to remove the divergencies one needs local counterterms which
are at least linear in the sources for the composite operators
\cite{I-Z}.  To remove the divergencies in all Green functions
one needs, in general case, a multitude of counterterms, each
containing the certain number of the sources or their
derivatives. The only bound at this stage is the requirement of
the Lorentz and parity invariance and that the dimension of the
counterterms should be equal to the dimension of a lagrangian
\cite{Shore}. In virtue of the chiral invariance the number of
the counterterms is highly limited.  One can show that only two
nontrivial counterterms, involving composite operators, are
needed.
\footnote{ We neglect here the renormalizations of the
fundamental fields of quarks and gluons, which should be
performed independently applying the standard technique
\cite{I-Z}.}
They are
\begin{equation}
(Z-1)\,\left(A^{0\mu} - \partial^{\mu}\Theta\right) J^{0}_{\mu 5}, \qquad
(Z_m-1) \sum_{a=0,1,\dots8}\left(-S^a\,J^a \! - \! P^a\,J^a_5\right),
\end{equation}
where $J^{0}_{\mu 5}$ is the singlet axial quark current and
$J^{a}$, $J^{a}_{5}$ are the scalar and pseudoscalar ones,
\begin{equation}
J^{0}_{\mu 5} = \bar q \gamma _{\mu } \gamma _{5} (\lambda ^{0}/2)q, \quad
J^{a} = \bar q(\lambda ^{a}/2)q, \quad
J^{a}_{5} = i\bar q\gamma _{5}(\lambda ^{a}/2)q.
\end{equation}
Both counterterms in (7) are chiral-invariant and of dimension
four. Note, the first counter\-term is chiral-invariant owing to
the derivative, acting on $\Theta$. In fact, it is rather
general result that $\Theta$ may contribute into a
chiral-invariant expression only with the derivative operator.
Hence, in virtue of dimensional reasons, the first counterterm
in (7) is the only one which depends on $\Theta$ and satisfies
the above conditions \cite{Shore}.  The renormalization constant
$Z$ for this counterterm was calculated in \cite{E-T}. The
second counterterm in (7) is like the mass-term one. Its
renormalization constant is independent of the quark flavours in
the mass-independent scheme.

The rest of counterterms, which are of the contact type (not
depending on the operators of the theory), are constructed from
invariant combinations of the sources and their derivatives:
\begin{eqnarray}
& &(F^R_{\mu\nu})^2, \quad (F^L_{\mu\nu})^2,                       \quad
[\partial^{\mu}(\partial_{\mu}\Theta - A^0_{\mu})]^2,              \nonumber\\
& &\partial^{\mu}(\partial_{\mu}\Theta - A^0_{\mu})\times(S^2+P^2),\quad
(S^2+P^2)^2.
\end{eqnarray}
Here the summation over omitted indexes is implied.  Notice,
according to the classification in Refs. \cite{G-L-1,G-L-2} all
these combinations have the common property to belong to order
$p^4$ or higher in the chiral dimension. Therefore, these
counterterms play no role in the generalization functional of
order $p^2$.

Usually, the introduction of counterterms (7) is interpreted as
a requirement of the multiplicative renormalization for
composite operators $J^0_{\mu}$, $J^a$, $J^a_5$ and of
nontrivial renormalization for the operator of the axial gluon
anomaly $Q$:
\begin{eqnarray}
& & J^{0}_{\mu 5\>R} = Z\>J^0_{\mu 5}, \quad
J^{a}_{\,R} =  Z_m\>J^{a},     \quad
J^{a}_{5\>R} = Z_m\>J^{a}_{5},    		     \nonumber \\
& & Q_{R} = Q - (1 - Z)\>\partial^{\mu}J^0_{\mu 5}.
\end{eqnarray}
Here the subscript {\small$R$} indicates the renormalized
operators. (Note, the last formula describes the quark-gluon
mixing due to the UV renormalization.) The rest of composite
operators introduced in (2) remains invariant. The renormalized
lagrangian may be obtained in the approach as a result of
substitution into the initial lagrangian (2) of the renormalized
operators instead of the bare ones and adding the contact
counterterms.

There is also an alternative way to describe counterterms, based
on the formal transformation of the sources \cite{Shore}. This
approach, the most convenient in the framework of generating
functional, is described as the following substitutions in
lagrangian (2):
\begin{eqnarray}
& & S^a \> = \> Z_m\>S^{a}_{\,R}, \>\quad
P^a \> = \> Z_m\>P^{a}_{\,R}, \>\quad  \Theta = \Theta_R, \nonumber\\
& & A^0_{\mu} = Z\,A^0_{\mu\,R} + (1-Z)\>\partial_{\mu}\Theta_{R}.
\end{eqnarray}
Notice, due to (11), the expression $\partial_{\mu}\Theta -
A^0_{\mu}$ is transformed multiplicatively:
\begin{equation}
\partial_{\mu}\Theta - A^0_{\mu} = Z\,(\partial_{\mu}\Theta - A^0_{\mu})_R.
\end{equation}
In formulae (11) and (12) the quantities, provided with
subscript {\small$R$}, are interpreted as the renormalized
sources. Contact counterterms (9) appear in this approach as a
result of nonlinear renormalization of the auxiliary source of
dimension four which should be added into the initial lagrangian
with the unit operator \cite{Shore}.  As a result of
substitutions (11) the generation functional
$W(S,P,\Theta,A^0,\dots)$ of the unrenormalized theory becomes
the generation functional for renormalized Green functions:
\begin{eqnarray}
W(S,P,\Theta,A^0_{\mu},\dots) & = &
W(Z_mS_R,\,Z_mP_R,\,\Theta_R,\,Z A^0_{\mu\,R}\!+\!
(1\!-\!Z)\partial_{\mu}\Theta_R,\dots)
		\nonumber\\
& \equiv & W_R(S_R,P_R,\Theta_R,A^0_{\mu\,R},\dots).
\end{eqnarray}
{}From (13) one can deduce the property of the renorm-invariance
of the generation functional written in terms of the
renormalized sources.

\section{The effective chiral lagrangian}

Calculating in (1) the functional integral over the variables
corresponding to colour degrees of freedom and heavy hadrons
ones, one can obtain the representation for the generalization
functional in terms of the effective theory. In case when only
the lightest mesons of the Goldstone nature are not integrated
out, we get:
\begin{equation}
e^{iW(V,A,S,P,\Theta)} = \int {\cal D}\left[U\right]\,
e^{i\int d^4 x {\cal L}_{eff}(U;V,A,S,P,\Theta)},
\end{equation}
where ${\cal L}_{eff}$ is the effective chiral lagrangian. The
interpolating fields for the mesons are accumulated in (14) in
the unitary $3\times3$ matrix $U$, satisfying the condition
$\det U = 1$.  Under the action of the chiral group
$SU(3)_{L}\times SU(3)_{R}$ matrix $U$ transforms like
\begin{equation}
U \to \Omega_L \, U \, \Omega_R^{\dagger},
\end{equation}
while the flavour-singlet transformations $U(1)_{L}\times
U(1)_{R}$ do not affect $U$.  Usually, $U$ is represented in the
exponential parameterization
\begin{equation}
U = \exp\left(i\sum_{a=1,\dots8}\lambda^a\eta^a\,/\,F\right),
\end{equation}
where $\eta^a$ are the interpolating fields for the mesons.

According to the prescriptions of the chiral perturbation theory
\cite{G-L-1,G-L-2} lagrangian ${\cal L}_{eff}$ in (14) may be
represented in the form of the expansion in the derivatives of
fields and sources. In the leading order $p^2$ of the expansion
${\cal L}_{eff}$ is described by nonlinear $\sigma$-model in the
presence of external fields:
\begin{eqnarray}
{\cal L}^{(2)}_{eff} & = & \frac{F^2}{4}\,
{\rm tr}\!\left(\nabla_{\mu}U\nabla^{\mu}U^{\dagger} +
\chi U^{\dagger} + \chi^{\dagger}U\right) +
H\,(\partial_{\mu}\Theta - A^0_{\mu})^2,		\\
\nabla_{\mu} U & = &
\partial_{\mu} U - i\widetilde{L}_{\mu}U +iU\widetilde{R}_{\mu}, \qquad
\chi = 2B(S+iP)\,e^{i\lambda^0\Theta}.
\end{eqnarray}
Here the tildes mean that the singlet sources $L^0_{\mu}$ and
$R^0_{\mu}$ are not taken into account in the definition of
$\nabla_{\mu} U$.  Parameter $F$ in (16), (17) stands for the
universal decay constant for the octet of mesons.  Parameter $B$
is connected with condensate of quarks.  $H$ describes the
contact term of the singlet sources.  One may associate $H$ with
the low-energy asymptotic of the propagator for the singlet
axial quark current.

Since quantum loops do not contribute into the effective theory
in the leading order, the generating functional in the
approximation is representable as
\begin{equation}
W^{(2)}(V,A,S,P,\Theta) = \int d^4 x {\cal L}^{(2)}_{eff}(U;V,A,S,P,\Theta),
\end{equation}
where $U$ is the solution to the classical equations of motion
in the presence of the sources. Due to (19), Green functions for
composite operators in QCD may be evaluated in terms of the
effective theory.

In virtue of (13) and (19), the QCD-inspired renorm-group
properties of the effective theory are reduced to the
requirement of the renorm-invariance of ${\cal L}^{(2)}_{eff}$.
In order to provide this property it suffices to demand the
renorm-invariance of $U$ and $F$ and the following
transformation rules for the constants $B$ and $H$:
\begin{equation}
B \, \to \, B_R\,=\,Z_mB,   \qquad  H \, \to \, H_R\,=\,Z^{2}H.
\end{equation}
The property of the renorm-invariance of $F$ and properties (20)
for $B$ and $H$ may be verified directly, keeping in mind the
above QCD descriptions for these constants. Therefore, the
requirement of the renorm-invariance of ${\cal L}^{(2)}_{eff}$
is equivalent to that of the renorm-invariance of the
interpolating fields $\eta^a$.

Let us now consider the generalization that involves the ninth
field $\eta^0$, responsible for the singlet member of the nonet
of non-excited quarkic states.  (One should retain the
$\eta^0$-integration in (14) in this case.) The general way of
the including of $\eta^0$ was outlined earlier in \cite{G-L-2}.
A possibility to determine $\eta^0$ as the very field of the
non-excited singlet quarkic state was based on the exclusive
property of this field to transform under the action of the full
chiral group $U(3)_{L}\times U(3)_{R}$ through adding a term
only, which is proportional to the parameter for
$U(1)_{A}$-transformation \cite{W}:
\begin{equation}
\eta^0 \, \to \, \eta^0 + F^0\omega^0_5.
\end{equation}
Thus, $\eta^0 + F^0\Theta$ remains completely chiral-invariant.

The quantity $F^0$ in (21) is a new parameter of the dimension
of mass. Its value, obviously, depends on the normalization of
$\eta^0$.  Usually, it is assumed that $F^0$ may be attributed
to the decay constant for the singlet axial quark current.
However, the latter is not renorm-invariant in view of (10).
Therefore, the assumption cannot be combined with the condition
of the renorm-invariance of $\eta^0$, which would be highly
desirable, especially, taking in mind the analogy with the
previous case of the octet fields. Thus, it is a question
whether it is really possible to introduce $\eta^0$ to be
renorm-invariant object. In case of the positive answer the next
question is what is the meaning of $F^0$ in terms of QCD.

Adjourning, temporarily, the discussion of these questions let
us consider, following \cite{G-L-2}, the most general form of
the chiral-invariant lagrangian involving $\eta^0$ up to and
including order $p^2$:
\begin{equation}
{\cal L}^{(2)}_{eff} = {\cal L}_0 +
\upsilon_1\,{\rm tr}\!\left(\nabla_{\mu}U\nabla^{\mu}U^{\dagger}\right) +
{\rm tr}\!\left(M\upsilon_2^{\ast} \Sigma^{\dagger} +
M^{\dagger}\upsilon_2 \Sigma \right).
\end{equation}
Here ${\cal L}_0$ stands for the lagrangian for $\eta^0$,
containing no contributions of matrix $U$.  In the third term,
conventionally called as ``mass'' one, the $\Sigma$ is the
matrix for the nonet of PS fields:
\begin{equation}
\Sigma = U\,e^{i\lambda^0\eta^0\,/\,F^0}.
\end{equation}
Notice, $\Sigma$ involves $\eta^0$ divided by $F^0$, so, it
transforms like $\Sigma \to \Omega_L \, \Sigma \,
\Omega_R^{\dagger}$ when $\Omega_{L,R} \in U(3)_{L,R}$. The
second, ``kinetic'', term in r.h.s. of (22) may be written in
terms of $\Sigma$, too. For this purpose one has to change the
definition of the covariant derivative allowing contributions of
the singlet sources in (18). However, in view of (11), this way
does not seem to be reasonable, because it is hard to define the
renorm-group properties of the theory in this case.  Therefore,
having in mind that the difference between the two ``kinetic''
terms does not depend on $U$ (and, hence, may be incorporated
into ${\cal L}_0$), let us keep the proposed above variant,
which is free from the dependence on the dangerous singlet
source $A^0_{\mu}$.

The essential moment, distinguishing (22) from (17), is the
presence of invariant functions $\upsilon_{1,2}$ and ${\cal
L}_0$, containing the dependence of the theory on $\eta^0$ and
the singlet sources. The only restriction on these functions is
that $\upsilon_1$ and ${\cal L}_0$ must be real and even,
whereas $\upsilon_2$ may be complex and $\upsilon_2^{\ast}(\vec
\alpha) = \upsilon_{2}(-\vec \alpha)$, where $\vec \alpha$
stands for the arguments \cite{G-L-2}. Since there are three
only invariant combinations of $\eta^0$, the singlet sources,
and their derivatives, we have in general case:
\begin{equation}
{\cal L}_0 = {\cal L}_0(\eta^0 \! + \! F^0\Theta,
\nabla_{\mu}\eta^0,\nabla_{\mu}\Theta), 	\quad
\upsilon_{i} = \upsilon_{i}(\eta^0 \! + \! F^0\Theta,
\nabla_{\mu}\eta^0,\nabla_{\mu}\Theta),
\end{equation}
where
\begin{equation}
\nabla_{\mu}\eta^{0} = \partial_{\mu}(\eta^{0} + F^0 \Theta), \quad
\nabla_{\mu}\Theta = h\,(\partial_{\mu}\Theta - A^{0}_{\mu}).
\end{equation}
Here in the definition of $\nabla_{\mu}\Theta$, the multiplier
$h$ is of the dimension of mass. Its role here is to equate the
dimensions of the both covariant derivatives.

Note, that assuming the dependence on the covariant derivatives
in $\upsilon_i$ and allowing for more then quadratic dependence
on the derivatives in ${\cal L}_0$, we have essentially diverged
from \cite{G-L-2}, where $\upsilon_i$ were considered as
derivative-free potentials and ${\cal L}_0$ did not more then
quadratic in the derivatives. The reason for the assumptions of
Ref. \cite{G-L-2} was that $\partial_{\mu}\eta^0$,
$\partial_{\mu}\Theta$ and $A^0_{\mu}$ have the chiral dimension
of $p^1$, which might be established starting from the equation
of motion for $\eta^0$ provided that the mass parameter for
$\eta^0$ is assigned to order $p^2$ in the chiral dimension.
The reason for the assignment was that in the chiral limit the
mass of $\eta^0$ tends to zero at large $N_c$ \cite{W}, so, it
may be made so small as needed. However, because the large-$N_c$
argumentation can lead, in fact, to serious consequences for the
chiral perturbation theory, we shall not resort to it in this
and next sections, where rather general properties of the
effective theory are discussed.

Indeed, thinking $\eta^0$ to be a chiral field (which is the
exact result in the limit of large $N_c$), and assuming ${\cal
L}_0$ to involve a derivative-free self-interaction of $\eta^0$,
then there are vertices of order $p^0$ in the chiral effective
theory. Therefore, defining the generalization functional at
order $p^2$ one should take into account the multiloop chiral
contributions and, moreover, the contributions of the higher
dimensions like $O(p^4)$, etc. This fact follows immediately
the formula for the overall chiral dimension for a connected
diagram with $L$ chiral loops and $N_d$ vertices of order $p^d$
($d=0,2,\dots)$:
\begin{equation}
D = 2 L + 2 + \sum_{d} N_d (d-2).
\end{equation}
The above result frustrates validity of the chiral perturbation
theory. However, the situation would not occur if $\eta^0$ had a
finite (non-vanish) mass, because in this case evaluating the
chiral dimension for a connected diagram one can think the mass
of $\eta^0$ as an effectively large parameter suppressing the
chiral contributions from $\eta^0$. Note, it does not mean that
one should no longer take into account the multiloop
contributions of $\eta^0$. It means only that in the case they
do not contribute in $D$ and, also, that there is no need
take into account the higher-dimensional vertices. That salvages
the chiral perturbation theory. The pay for the salvation is
loss of the simple representation for the generalization
functional, like (19), because in the case the leading order of
the chiral expansion does not coincide with quasi-classical
approximation.

The analysis of the renorm-group behaviour of the effective theory
inspiring by QCD renormalization presents no insuperable
problems after the above consideration. The principal moment is
to prove the renorm-invariance of the parameter $F^0$. To
provide for this property one should require the function
$\upsilon_{2}$ to transform like $B$ in (20) and the parameter
$h$ in such a way to ensure the renorm-invariance of the
covariant derivative $\nabla_{\mu}\Theta$. Owing to (12), this
requirement is true if $h^2$ transformed like $H$ in (20), i.e.
\begin{equation}
h \to h_R = Z\,h.
\end{equation}
Since the theory depends on $A^0_{\mu}$ through the covariant
derivative $\nabla_{\mu} \Theta$ only, the parameter $h$ may be
attached to the decay constant for the singlet axial quark
current. In fact, owing to (10), this observation proves
property (27) and, so, the statement that $F^0$ is
renorm-invariant. Consequently, the interpolating field $\eta^0$
is renorm-invariant, as well. The meaning of the constant $F^0$
will be discussed below.

The generalization of the results for the case when an
additional singlet interpolating field (fields) $\eta^G$ is
involved may be performed by analogy. The essential difference
between $\eta^G$ and $\eta^0$ is that $\eta^G$ is described as a
complete singlet, i.e. it is not affected by any chiral
transformation, including $U(1)_A$.  Therefore to introduce
$\eta^G$ into the effective theory one should simply include the
dependence on $\eta^G$ and its derivatives into the invariant
functions $\upsilon_{i}$ and extra lagrangian term ${\cal L}_0$.
Then, assuming the renorm-invariance for $\eta^G$, no properties
of the theory are changed through the including. The general
question now is what is the nature of $\eta^G$. Its possible
interpretation is either a glueball or an excitation over
$\eta^0$. (We do not consider here the heavy quark and
multi-quark contributions). The difference between both cases
may be revealed through the study of the relationship between
the quarkic and gluonic composite operators of QCD and currents
of the effective theory. Another way is to study the typical
features in dependence of ${\cal L}_0$ and $\upsilon_i$ on
$\eta^G$ in both cases. The research of these questions is the
aim of the next sections.

\section{The currents of the effective theory}

The natural way to introduce the currents of the effective
theory, related to the composite operators of QCD, is through the
variational derivatives of the action of the effective theory on
the very sources for the composite operators. For instance, the
scalar and pseudoscalar currents are defined as the first
derivatives on the sources $S^a$ and $P^a$. So, in the context
of lagrangian (22) we have
\begin{eqnarray}
{\cal J}^a & = & -\> \delta {\cal L}^{(2)}_{eff}/\delta S^a =
		{\rm tr}\left\{\frac{\lambda^a}{2}\left(
		-\upsilon_2 \Sigma - \upsilon_2^{\ast}\Sigma^{\dagger}
		\right)\right\},					\\
{\cal J}^a_5 & = & -\> \delta {\cal L}^{(2)}_{eff}/\delta P^a =
		{\rm tr}\left\{\frac{\lambda^a}{2}\left(
		i\upsilon_2 \Sigma - i\upsilon_2^{\ast}\Sigma^{\dagger}
		\right)\right\}.
\end{eqnarray}
(For brevity, we use the one and the same symbol for the action
and for the lagrangian.)

In virtue of lagrangian (22) depends linearly on $S^a$ and
$P^a$, the very currents ${\cal J}^a$ and ${\cal J}^a_5$ possess
the property of independence on the sources themselves.  Since
this property and equality $W^{(2)}_{QCD} = W^{(2)}_{eff}$, we
have the following relations between the matrix elements in
QCD and ones in the effective theory made of the identical sets
of operators $J^a$, $J^a_5$ and ${\cal J}^a$, ${\cal J}^a_5$:
\begin{equation}
<a|J^{a_1} \dots J^{a_n}
\dots J^{a_{n+m}}_{5}|b>_{_{QCD}}
\> = \>\,
<a|{\cal J}^{a_1} \dots {\cal J}^{a_n}
\dots {\cal J}^{a_{n+m}}_{5}|b>_
{_{\!\!\!{\hbox{\scriptsize $eff$}}}}.
\end{equation}
Here $<a|$ and $|b>$ stands for the vacuum state or any other
states described by the effective theory ($\pi,K,\eta,\eta'\dots)$.

Relation (30) means that in $p^2$-approximation the composite
operators $J^a$, $J^a_{5}$, operating in QCD (in the indicated
above space of states), act identically to the operators ${\cal
J}^a$, ${\cal J}^a_5$, operating in the effective theory.  The
direct consequence from this observation is the requirement that
both sets of operators should have identical chiral-symmetry
properties at fixed sources. In case of the transformations
$SU(3)_L \times SU(3)_R$ and $U(1)_V$ this requirement is
fulfilled automatically in view of the transformation rule (15)
and the property that $\eta^0$ is the exact singlet under these
transformations. In case of the axial-singlet transformation
$U(1)_A$ the requirement leads to the nontrivial consequence.
Indeed, as directly follows from exact expressions (28), (29),
and due to $U(1)_A$-transformation properties for $\Sigma$, the
transformation rules required for ${\cal J}^a$, ${\cal J}^a_5$
are only fulfilled when $\upsilon_2$ is not changed under the
transformation.  At the fixed sources the latter only can take
place when $\upsilon_2$ has no dependence on the field $\eta^0$
and its derivatives. In view of (24), this means that
$\upsilon_2$ does not depend on its allowing arguments $\eta^0 +
F^0\Theta$ and $\nabla_{\mu}\eta^0$. Thus, $\upsilon_2$ may
depend on $\eta^G$, $\partial_{\mu} \eta^G$ and $\nabla_{\mu}
\Theta$ only. The equivalent proof of this result, based on the
analysis of permutation relations between the currents and the
generator of $U(1)_A$-transformation, is given in Appendix.

The restriction obtained on the function $\upsilon_2$ allows one
to establish the important corollary concerning the properties
of the singlet-field lagrangian ${\cal L}_0$. The idea is to
inspect if the dependence on $\eta^0$ will appear in
$\upsilon_{2}$ when $\eta^G$ is integrated out from the theory.
It is easy to show that in the case when $\upsilon_2$ involves
$\eta^G$ or its derivatives, the dependence cannot appear if
${\cal L}_0$ admits of no interaction between $\eta^0$ and
$\eta^G$. (In this case $\eta^0$ does not contribute into the
equation of motion for $\eta^G$ in the leading order $p^0$ of
the lagrangian (22). The contributions into $\upsilon_2$, going
through the dependence on $\eta^0$ and $\eta^G$ in the
$p^2$-terms of the lagrangian, are irrelevant here, because in
the end they contribute beyond the order $p^2$ of the
lagrangian.) In this case ${\cal L}_0$ is representable as the
sum of two independent lagrangians, one for $\eta^0$ and another
for $\eta^G$:
\begin{equation}
{\cal L}_0 = {\cal L}_{\eta^0}(\eta^0 \! + \! F^0\Theta,\,
			\nabla_{\mu}\eta^0,\,\nabla_{\mu}\Theta) +
			{\cal L}_{\eta^G}(\eta^G,\,
			\partial_{\mu}\eta^G,\,\nabla_{\mu}\Theta).
\end{equation}
The second case is when $\upsilon_2$ involves neither $\eta^G$
nor $\partial_{\mu}\eta^G$. In this case ${\cal L}_0$ may well
contain an additional term describing the interaction between
$\eta^0$ and $\eta^G$.

The first case considered above ($\upsilon_2$ depends on
$\eta^G$ or its derivatives) means that $\eta^0$ and $\eta^G$
can interact with each other without the octet fields
contribution through the ``mass'' term only. When the quark
masses vanish and the sources are turned-off they cannot
interact at all without the octet fields which become the
Goldstone bosons in the limit. Such behaviour is typical for
objects one of which is an excited state, because the latter in
the chiral limit seems cannot enter into the strong interaction
with any object without the emission of the Goldstone bosons. On
this ground one may conclude that the most probable
interpretation for $\eta^G$ in this case is an excitation over
$\eta^0$, i.e.  the quark excitation or a hybrid state, in
dependence of the type of the degrees of freedom being excited
in $\eta^0$.  In the second case ($\upsilon_2$ does not depend
on $\eta^G$ and its derivatives) both fields, $\eta^0$ and
$\eta^G$, can well interact with each other without the
Goldstones when quark masses and sources are turned-off. This
picture of interaction is typical for non-excited states.  Since
there is not another state neighboring in energy, one may
consider $\eta^G$ as a PS-glueball in this case.

Let us now consider the singlet currents ${\cal Q}$ and ${\cal
J}^0_{\mu 5}$ related to the QCD composite operators $Q$ and
$J^0_{\mu\,5}$:
\begin{equation}
{\cal Q} =  \delta{\cal L}^{(2)}_{eff}/\delta \Theta, \quad
{\cal J}^0_{\mu 5} = \delta{\cal L}^{(2)}_{eff}/\delta A^0_{\mu}.
\end{equation}
In contrast to the above case, these currents act not
identically to the operators $Q$ and $J^0_{\mu\,5}$ determined
in the space of states of the effective theory. (The same
property may be established as well for the octet-vector and
octet-axial quark currents.) That fact follows immediately from
the nonlinear character of the dependence of lagrangian (22) on
the corresponding sources.  However, any relation linear in the
composite operators should take place as well in the effective
theory.  For instance, it is easy to show by the straightforward
calculation that following the QCD renormalization the singlet
currents ${\cal Q}$ and ${\cal J}^0_{\mu 5}$, although being
made of the renorm-invariant interpolating fields, transform
like:
\begin{equation}
{\cal J}^{0}_{\mu 5\>R} = Z\>{\cal J}^0_{\mu 5}, \quad
{\cal Q}_{R} = {\cal Q} - (1 - Z)\>\partial^{\mu}{\cal J}^0_{\mu 5}.
\end{equation}
It can be shown also that ${\cal Q}$ and ${\cal J}^0_{\mu 5}$
satisfy the Ward identity (on the equations of motion for the
interpolating fields) which coincides with the anomalous Ward
identity in QCD for $Q$ and $J^0_{\mu 5}$. The relations of the
kind of (30) are also true provided the operators $Q$ or
$J^0_{\mu 5}$ and ${\cal Q}$ or ${\cal J}^0_{\mu 5}$ were
inserted only once.  In particular, the following relations take
place
\begin{equation}
<0|Q|\eta^{0,G}\!> \> = \> <0|{\cal Q}|\eta^{0,G}\!>, \quad
<0|J^0_{\mu 5}|\eta^{0,G}\!> \> = \> <0|{\cal J}^0_{\mu 5}|\eta^{0,G}\!>.
\end{equation}
{}From the above properties it is natural to regard ${\cal Q}$ and
${\cal J}^0_{\mu 5}$ as the effective-theory analogs to the
gluonic operator $Q$ and axial singlet quark current $J^0_{\mu
5}$ of QCD. The analogy would be strong when only the linear
correlations were considered. Notice, equalities (34) allow one
to consider the properties of $\eta^0$ and $\eta^G$, initially
introduced in the effective theory, to be dependent on the
behaviour of the matrix elements $<0|Q|\eta^{0,G}\!>$ and
$<0|J^0_{\mu 5}|\eta^{0,G}\!>$ defined in QCD. This property
will be exploited below for further discussion of the
differences between the glueball and excitation over $\eta^0$.

\section{The spectrum of the effective theory}

To investigate the spectrum of the effective theory it is
reasonable to combine the chiral expansion with quasi-classical
approximation. Then, lagrangian (22) may be regarded as the
effective action for the mesons in the presence of the external
fields, which are the sources for the QCD composite operators,
too. Applying the combined approximation we are able also to
make use the large-$N_c$ approximation, which can assist us to
recognize the nature of the extra singlet field $\eta^G$.

At first, let us study the lagrangian ${\cal L}_0$ in the
quadratic approximation on the fields and sources. Starting from
the most general expression for ${\cal L}_0$ provided with the
symmetry required, we have
\begin{eqnarray}
{\cal L}_0           & = &
\frac{1}{2}\alpha_1 (\nabla_{\mu}\eta^0) (\nabla^{\mu}\eta^0) +
\frac{1}{2}\alpha_2 (\partial_{\mu}\eta^G)(\partial^{\mu}\eta^G) +
\frac{1}{2}\alpha_3 (\nabla_{\mu}\Theta)(\nabla^{\mu}\Theta)
\nonumber \\	     &   &
+\, \alpha_4 (\nabla_{\mu}\eta^0)(\partial^{\mu}\eta^G) +
\alpha_5 (\nabla_{\mu}\eta^0)(\nabla^{\mu}\Theta) +
\alpha_6 (\partial_{\mu}\eta^G)(\nabla^{\mu}\Theta)
\nonumber \\	     &   &
-\,\frac{1}{2}\beta_1 (\eta^0 + F^0\Theta)^2
- \frac{1}{2}\beta_2 (\eta^G)^2
- \beta_3 (\eta^0 + F^0\Theta)\eta^G.
\end{eqnarray}
Here $\alpha_i$ and $\beta_i$ are the constants.  Some of them
may be removed or fixed through more special consideration. In
this way, the constant $\alpha_1$ may be absorbed by the
normalization for the field $\eta^{0}$ and parameter $F^0$. So,
without loss of generality we may set $\alpha_1 = 1$ providing
for the canonical normalization for $\eta^0$.  The fourth term
in (35) may be removed out by the linear transformation
\begin{equation}
\eta^0 \to \eta^0 - \alpha_4 \, \eta^G,
\end{equation}
diagonalizing the kinetic terms in (35). Owing to the
transformation (36) does not run counter to (21), it is allowed
for $\eta^0$.  Generally, any transformation of the kind
describes the uncertainty in definition of $\eta^0$, which may
be associated with the indeterminate contributions of gluons and
other types of contributions into the singlet state of quarkic
nature. The very transformation (36) fixes the uncertainty in
the context of the effective theory, and then $\eta^0$ becomes
the interpolating field for observable state.

Performing transformation (36) and fixing the canonical
normalization for $\eta^G$, let us rewrite ${\cal L}_0$ in the
form
\begin{eqnarray}
{\cal L}_0    & = &
\frac{1}{2}\,(\nabla_{\mu}\eta^0) (\nabla^{\mu}\eta^0)
+ \frac{1}{2}\,(\partial_{\mu}\eta^G) (\partial^{\mu}\eta^G)
+ \frac{1}{2}\alpha_{\Theta} (\nabla_{\mu}\Theta) (\nabla^{\mu}\Theta)
\nonumber \\  &   &
+\, \alpha_0 (\nabla_{\mu}\eta^0) (\nabla^{\mu}\Theta)
+ \alpha_G (\partial_{\mu}\eta^G) (\nabla^{\mu}\Theta)
\nonumber \\  &   &
-\, \frac{1}{2} M^2_0 (\eta^0 + F^0\Theta)^2
- \frac{1}{2} M^2_G (\eta^G)^2 - q (\eta^0 + F^0\Theta)\eta^G.
\end{eqnarray}
Here $M_0$, $M_G$ are the mass parameters for the singlet
interpolating fields and $q$ is the parameter describing their
mixing.

Due to (32) and (37), the currents ${\cal J}^0_{\mu 5}$ and
${\cal Q}$ in the linear approximation on the fields and turned-off
sources may be represented as
\begin{eqnarray}
{\cal J}^{0}_{\mu 5} & = &
-\>h\,\partial_{\mu} (\alpha_0 \eta^0 + \alpha_G \eta^G), \\
{\cal Q} & = & -\>F^0 (M^2_0 + \partial^{2})\eta^0 -
F^0 q \eta^G - h\,\partial^{2} (\alpha_0 \eta^0 + \alpha_G \eta^G ) \simeq
\nonumber   \\       &   & \quad
h (\alpha_G M^2_G + \alpha_0 q) \eta^G +
h (\alpha_0 M^2_0 + \alpha_G q) \eta^0 + (mass).
\end{eqnarray}
Here symbol `$\simeq$' means that the equations of motion were
used. The last term in (39), designated like `$(mass)$', stands
for the contributions which are proportional to the current
quark masses.  Owing to (34), $h\alpha_0$ in (38) is equal to
the decay constant for the singlet axial quark current. Assuming
$\alpha_0$ to be absorbed by $h$ we may set $\alpha_0 = 1$.

Using (34) and (38), (39), one can determine the large-$N_c$
behaviour of the parameters of lagrangian ${\cal L}_0$ depending
on the nature of $\eta^G$. To this end, let us put the quark
masse to be vanish (the chiral limit) and consider the following
well-known formulae \cite{W}:
\begin{eqnarray}
<0|J^0_{\mu 5}|\mbox{glueb}> \> \sim 1,       &  &
<0|J^0_{\mu 5}|\mbox{quark}> \> \sim N_c^{1/2}, \quad \\
<0|\,Q\,|\mbox{glueb}> \>\, \sim \, 1,        &  &
<0|\,Q\,|\mbox{quark}> \> \sim N_c^{-1/2}.
\end{eqnarray}
{}From here and (38), (39), taking into consideration the quark
nature of $\eta^0$, and the property $M_G \sim 1$ postulated
independently on the nature of $\eta^G$, one can deduce that
$M^2_0 \sim N_c^{-1}$ and $h \sim N_c^{1/2}$, and also that
\begin{equation}
q \sim  N_c^{-1/2},  \quad     \alpha_G \sim N_c^{-1/2}
\end{equation}
for $\eta^G$ is a glueball, and
\begin{equation}
q \sim  N_c^{-1}, \quad \alpha_G \sim N_c^{-1}
\end{equation}
for $\eta^G$ is an excitation over $\eta^0$.

The spectrum of the effective theory in the chiral limit can be
determined by diagonalizing the mass terms in (37).  The final
result at large $N_c$  looks like
\begin{equation}
M^2_{\eta'_0} = (M^2_0 M^2_G - q^2)/M^2_G, \qquad
M^2_{\eta''_0} = M^2_G.
\end{equation}
Here the symbols $\eta'$ and $\eta''$ represent the observable
states having a certain value of mass (the subscript zero means
that the states are considered in the chiral limit). From the
first equality in (44) and (42), (43) one may deduce that
$M^2_{\eta'_0}\sim N_c^{-1}$, and that in the leading order in
$N_c^{-1}$ the mixing parameter $q$ can contribute into
$M^2_{\eta'_0}$ if $\eta^G$ is the glueball only. On the
contrary, when $\eta^G$ is the excitation over $\eta^0$, then
$q$ can contribute in the next-to-leading order only (if it does
not equal identically zero). Therefore it must be eliminated
from the theory in the case in $p^2$-approximation when one
equates the order of magnitude of $O(N_c^{-1})$ to $O(p^2)$.
Note, the similar result has been obtained in the preceding
section on the ground of rather general consideration.

Now let us introduce the ``mass'' term of the effective
lagrangian. Putting
\begin{equation}
\upsilon_2 = \frac{1}{2}BF^2 \left(1 + ib\,\eta^G/F^0 + \dots\right),
\end{equation}
where $b$ is a new constant responsible for
$\eta^G\!$-dependence in $\upsilon_2$, we get from (29) in the
linear approximation:
\begin{eqnarray}
{\cal J}^{a\not=0}_5 \!\!\!\! & = & - BF^2\,\left(\eta^a/F\right), \\
{\cal J}^0_5 & = & - BF^2\left(\eta^0 + b \eta^G\right)/F^0.
\end{eqnarray}
Owing to the similar to (40) formula for the large-$N_c$
behaviour of the PS quark currents, and since $F \sim
N_c^{1/2}$, it follows from (47) that $F^0 \sim N_c^{1/2}$.
Then, depending on the nature of $\eta^G$, the parameter $b$ in
(47) behaves at large $N_c$ as
\begin{equation}
b \sim \cases{N_c^{-1/2}, & {\rm for $\eta^G$ is a glueball}\cr
	      1, & {\rm for $\eta^G$ is an excitation over $\eta^0$.}\cr}
\end{equation}
We see that in the first case the parameter $b$ is suppressed by
large $N_c$. Therefore, when determining the spectrum in
$p^2$-approximation combined with large $N_c$, one should
eliminate $b$, because it contributes through the ``mass'' term
which is already suppressed. In the second case of (48)
parameter $b$ may well contribute in $p^2$-approximation.

The consequence from (47) and (34), which is of the great
importance, is the formula for the parameter $F^0$, representing
it in terms of QCD variables:
\begin{equation}
F^0 = \frac{<\bar u u>_0}{<0|J^0_5|\eta^0>}.
\end{equation}
Here the equality  $BF^2 = -<\bar u u>_0$ \cite{G-L-2} has been
exploited where $<\bar u u>_0$ is the chiral quark condensate
($<\bar u u>_0 = <\bar d d>_0 = <\bar s s>_0$). Note, to within
designations, (49) is equivalent to the result obtained earlier
in the framework of PCAC \cite{S-V}.

Turning-on the quark masses one may obtain the mass matrix for
the observable states. When $m_u = m_d \not= m_s$ it describes
the $\eta^8 - \eta^0 - \eta^G$ mixing. In this very basis the
squared mass matrix is
\begin{equation}
{\cal M}^2 = \left( \begin{array}{ccc}
\frac{1}{3}(4 M_K^2\!-\!M_{\pi}^2) &
	\frac{2\sqrt{2}}{3}\xi (M_{\pi}^2 - M_K^2) &
	\frac{\sqrt{2}}{3}\,b\,\xi (M_{\pi}^2 - M_K^2)         \\[2mm]
	& M_0^2 + \frac{1}{3}\,\xi^2 (2 M_K^2\!+\!M_{\pi}^2)
	& q + \frac{1}{3}\,b\,\xi^2(2 M_K^2\!+\!M_{\pi}^2)     \\[2mm]
\mbox{ symm.}		& & M_G^2			       \\
\end{array} \right)
\end{equation}
Here $M_{\pi}$ and $M_K$ are the pion and kaon masses, $\xi =
F/F^0$. If $\eta^G$ is an excitation over $\eta^0$, then one
should put $q = 0$ in (50). When $\eta^G$ is a glueball, then $b
= 0$.  Notice, in the second case with $\xi = 1$ the matrix
${\cal M}^2$ is equivalent to the Kawai matrix \cite{K} with the
two parameters (not counting $M_G^2$) instead of three ones in
\cite{K}.

The eigenvalues of the squared mass matrix (50), which are the
squared masses of eigenstates $\eta$, $\eta'$, $\eta''$
resulting from the mixing of $\eta^8$, $\eta^0$, $\eta^G$, may
be evaluated by fitting the data for the radiative decays $P \to
\gamma\gamma$, $P \to V\gamma$ and $V \to P\gamma$, where $P =
\eta,\eta'$, $V = \omega,\rho$.  In the effective chiral theory
these decays, violating the internal parity, are described by
Wess-Zumino term (see, e.g., \cite{Bando,B}). In the approach
considered here this term should be constructed over the field
matrix $\Sigma$, involving $\eta^0$ divided by $F^0$.  Omitting
the tedious and rather standard calculation, let us present the
final result for the masses of $\eta$, $\eta'$, $\eta''$ in the
case when $\eta^G$ is the glueball:
\begin{eqnarray}
& & M_{\eta}^2 = (0.52 \pm 0.02\,\mbox{GeV})^2, \quad
M_{\eta'}^2 = (0.99 \pm 0.13\,\mbox{GeV})^2, \nonumber \\
& &M_{\eta''}^2 = (0.00 \pm 3.74)\,\mbox{GeV}^2 \, \leq \>
(1.94\,\mbox{GeV})^2.
\end{eqnarray}
The errors in (51) are the consequences of the 20\%-errors,
assumed in the mass formulae for $\eta$, $\eta'$, $\eta''$, and
the errors of the experimental data, which were put into the
fitting procedure. On definition, the quantity $M^2_{\eta''}$
was assumed to be non-negative through the fitting.

As one can see from (51) the estimate for $\eta''$ is very rough
to make any conclusion about for what real state the $\eta''$
stands. To solve this problem one needs study in detail the
decays of $\eta''$, which is beyond the framework of the present
work.  Notice only, that for the study it is important to know
\cite{G} whether the parameter $b$ in (45) is really equal
to zero when $\eta^G$ is the glueball.  According to section 4
it should exactly equal zero if the mixing parameter $q$ does
not.  However, for the glueball there is not strong restriction
that $q$ should differ zero at any price. Therefore, it is
desirable to get the quantitative estimate.  Unfortunately, the
result of the above fitting, which is $q = 0.1 \pm 1.0$, permits
no certain conclusion. Note, that the another result $\xi = 0.92
\pm 0.12$ allows one to conclude that the normalization constant
$F^0$ for the singlet field $\eta^0$ is fitted with high
accuracy and that it coincides within errors with the universal
decay constant $F$ for the octet of mesons.

\section{Summary and discussion}

The present paper has shown that the UV renormalization, mixing
in QCD the quarkic and gluonic composite operators (which
generate the $\eta'$-meson and PS-glueball), in the effective
theory does not affect the mutual configuration of the
interpolating fields for quarkic and gluonic states.
Nevertheless, the QCD renormalization of the composite operators
may well be reproduced as the renormalization of the related
currents of the effective theory. The interpolating field
$\eta^0$ for the lowest singlet quarkic state may be introduced
into the effective theory being normalized on the very special
renorm-invariant constant. (It is determined as the ratio of the
chiral quark condensat to the normalization constant for the
singlet pseudoscalar quark current. However, its value coincides
within errors with the universal octet decay constant
$F_{\pi}$.) The interpolating field for the lowest gluonic state
may be involved so that it saturates at large $N_c$ the gluonic
current of the effective theory.

The general way to involve singlet fields into the effective
chiral lagrangian is through the potentials describing the
``kinetic'' and ``mass'' terms of the lagrangian, and through
some extra terms which are the kinetic and mass terms of the
singlet fields themselves and their mutual- and
self-interaction. The present paper investigation has shown that
the interpolating field $\eta^0$ makes no contribution into the
potential of the ``mass'' term.  Consequently, another singlet
interpolating field, $\eta^G$, may only contribute into the
``mass'' term when there is not direct interaction between
$\eta^G$ and $\eta^0$ out of the ``mass'' term without the octet
fields contribution. The latter property is shown to be peculiar
for an excitation over $\eta^0$. On the contrary, PS-glueball
may well enter the interaction. In particular, it can mix with
$\eta^0$ in the chiral limit and, then, it makes no contribution
into the ``mass'' term of the effective chiral lagrangian.

The latter property results in serious consequences for the
decay modes of the PS-glueball. So, Ref. \cite{G} has shown that
with non-vanish parameter $b$, describing the contribution of
$\eta^G$ into the ``mass'' term of the lagrangian (in Ref.
\cite{G} this state is unreasonably identified with the
PS-glueball), the principal decay mode of $\eta^G$ is
predominantly $K\bar K \pi$.  In the case when $b$ vanishes,
this mode occurs through the mixing of $\eta^G$ with $\eta^0$
and $\eta^8$ only. If the mixing is large there may be the
copious decay. However, to get to know more on the question one
needs the additional study. The results obtained above may serve
as the first step in this trend.

\begin{flushleft}
{\large\bf Acknowledgements }
\end{flushleft}
The author is grateful to B.A.Arbuzov, V.E.Rochev and
F.V.Tkachov for helpful discussions. The work was supported in
part by RFFI grant No 94-02-03545-a.

\appendix
\section*{Appendix}
\noindent
The $U_A(1)$-transformation properties for the scalar and
pseudoscalar currents are determined by the permutation
relations
\begin{equation}
\left[\,{\cal O}^0_5,\,{\cal J}^a\right] = i\,d^{0ab}{\cal J}^b_5, \quad
\left[\,{\cal O}^0_5,\,{\cal J}^a_5\right] = -\,i\,d^{0ab}{\cal J}^b.
\end{equation}
Here $d^{0ab} = \sqrt{2/3}\>\delta^{ab}$, ${\cal O}^0_5$ is the
generator for $U_A(1)$-transformations. According to the
standard construction it equals the spatial integral over the
temporal component of the Noether current $\Im^0_{\mu 5}$. Owing
to (21), the latter may be represented as
\begin{equation}
\Im^0_{\mu 5} = F^0\,\frac{\partial{\cal L}^{(2)}_{eff}}
			{\partial(\partial_{\mu}\eta^0)}.
\end{equation}
It is not difficult to show that, due to (28) and (29), both
relations in (A.1) are equivalent to the single relation
\begin{equation}
\left[\,{\cal O}^0_5,\,\upsilon_2 e^{i\lambda^0 \eta^0 /F^0} \right] =
\lambda^0 \upsilon_2 e^{i\lambda^0 \eta^0 /F^0}.
\end{equation}

Let us now make use the fact that in view of (A.2) the temporal
component of $\Im^0_{\mu 5}$ coincides, up to the factor
$F^0$, with the canonical momentum, conjugated to $\eta^0$. From
here and in view of the canonical permutation relations for
$\eta^0$, one can deduce the following permutation relations:
\begin{equation}
\left[\,\Im^0_{0\, 5}({\rm \hbox{\bf x}}),\,
\eta^0({\rm \hbox{\bf y}}) \right] =
-\, i F^0 \delta({\rm \hbox{\bf x}} - {\rm \hbox{\bf y}}), \quad
\left[\,\Im^0_{0\, 5}({\rm \hbox{\bf x}}),\,
\partial_{n}\eta^0({\rm \hbox{\bf y}}) \right] =
i F^0 \partial_{n}\delta({\rm \hbox{\bf x}} - {\rm \hbox{\bf y}}).
\end{equation}
Here $n$ runs over the spatial values $n=1,2,3$. Thanks to
(A.4), one can write the permutation relation for any operator
$\Phi$, admitting a power decomposition:
\begin{equation}
\left[\,{\cal O}^0_{5},\,\Phi \right] =
-\, i F^0 \left[\frac{\partial \Phi}{\partial\eta^0} -
\partial_n \frac{\partial \Phi}{\partial(\partial_n\eta^0)} -
{\cal F}_0 \left(\frac{\partial \Phi}{\partial(\partial_{0}\eta^0)}\right)
\right].
\end{equation}
Here ${\cal F}_0$ is a functional satisfying the condition
${\cal F}_{0}(0) = 0$. (It is possible that ${\cal F}_0$
identically equals zero. In general case, ${\cal F}_0$ arises in
(A.5) because of the commutator of ${\cal O}^0_5$ with
$\partial_{0}\eta^0$ from $\Phi$.) Applying (A.5) to l.h.s of
(A.3) one gets the equation on $\upsilon_2$:
\begin{equation}
\frac{\partial \upsilon_2}{\partial\eta^0} -
\partial_n \frac{\partial \upsilon_2}{\partial(\partial_n\eta^0)} -
{\cal F}_0 \left(
\frac{\partial \upsilon_{2}}{\partial(\partial_{0}\eta^0)}\right) = 0.
\end{equation}

Due to Lorentz-invariance, (A.6) means that
$\partial\upsilon_2/\partial(\partial_{\mu}\eta^0) = 0$ and
$\partial \upsilon_2/\partial\eta^0$ $ = 0$, q.e.d. If second
and third terms in (A.6) taken together form a Lorentz-invariant
combination $\partial_{\mu}\partial\upsilon_2 /
\partial(\partial_{\mu}\eta^0)$, then (A.6) becomes simply
$\delta \upsilon_2/\delta \eta^0 = 0$. From here, owing to
arbitrariness of $\eta^0$, the same result follows.

\end{document}